\documentclass[11pt]{article}

\usepackage[a4paper,margin=1in]{geometry}
\usepackage[T1]{fontenc}
\usepackage[utf8]{inputenc}
\usepackage{lmodern}

\usepackage{amsmath,amssymb,amsthm,mathtools,bm}
\usepackage{microtype}

\usepackage[hidelinks]{hyperref}
\usepackage[nameinlink,noabbrev]{cleveref}

\usepackage{enumitem}
\usepackage{authblk}
\usepackage{setspace}
\usepackage{titlesec}
\usepackage{indentfirst}
\usepackage{cite}

\hypersetup{
  pdftitle={Asymptotically Weyl-invariant gravity: conditionally on-shell renormalizable in the ultraviolet},
  pdfauthor={Daniel Coumbe}
}

\titleformat{\section}{\large\bfseries}{\thesection}{0.7em}{}
\titleformat{\subsection}{\normalsize\bfseries}{\thesubsection}{0.6em}{}

\newcommand{\RPal}{\mathcal{R}}
\newcommand{\Ric}{\mathcal{R}_{\mu\nu}(\Gamma)}

\title{Ultraviolet Closure in Asymptotically Weyl-Invariant Gravity}
\author{Daniel Coumbe}
\affil{The Niels Bohr Institute, University of Copenhagen\\Blegdamsvej 17, DK-2100 Copenhagen \O, Denmark\\\texttt{DC@nrgym.dk}}
\date{}

\begin{document}
\maketitle

\begin{abstract}
We investigate asymptotically Weyl-invariant gravity (AWIG) in the Palatini formulation, defined by an exponent that interpolates between an Einstein-like infrared regime, where \(n\to1\), and a Weyl-invariant ultraviolet regime, where \(n\to2\). In four dimensions, we show that exact Weyl invariance within the minimal scalar Palatini \(f(\mathcal{R})\) sector uniquely selects the \(\mathcal{R}^2\) action as the ultraviolet endpoint. Interpreting the exponent as an effective curvature-dependent quantity, we constrain a phenomenological class of smooth autonomous flows compatible with the required infrared and ultraviolet endpoints. We then analyse the strict ultraviolet theory. On the regular branch \(\mathcal{R}\neq0\), and within a restricted torsionless, parity-even sector constructed purely from curvature, all admissible divergent on-shell counterterms reduce at arbitrary loop order to the original \(\sqrt{-g}\,\mathcal{R}^2\) term plus a topological Euler density. Although this conditional result does not establish full off-shell perturbative renormalizability, it provides nontrivial evidence for improved ultraviolet behaviour relative to Einstein gravity and identifies a concrete target for a future first-principles renormalization-group calculation.
\end{abstract}

\section{Introduction}

General relativity is extraordinarily successful in the infrared, but there are strong reasons to expect that it is not the final ultraviolet description of gravity. Classical solutions generically develop singularities at high curvature~\cite{Senovilla:2014gza}, and perturbative quantisation of the Einstein--Hilbert action yields a non-renormalizable theory~\cite{tHooft:1974bx,Goroff:1985th}. These difficulties suggest that Einstein gravity should be modified in the regime where it fails, namely in the ultraviolet, while still being recovered at low curvature. The question is then what principle, if any, should organise that ultraviolet modification.

A natural possibility is Weyl invariance, under which the metric transforms as $g_{\mu\nu}(x)\longrightarrow \widetilde{g}_{\mu\nu}(x)=\Omega^{2}(x)g_{\mu\nu}(x)$. The basic reason this is attractive is that Weyl rescalings leave the causal structure of spacetime unchanged. They preserve null cones, and hence the conformal structure determined by the metric, while changing only local scale. This matters because causality appears to capture many of the fundamental aspects of spacetime geometry. Results due to Malament~\cite{Malament1977} and to Hawking et al.~\cite{Hawking:1976fe} show that, under suitable conditions, causality determines almost all spacetime structure. These results are usually regarded as incomplete because the conformal factor remains undetermined. If the conformal factor becomes redundant in the ultraviolet, as Weyl invariance implies, then the principal obstruction preventing a complete causal description of spacetime would be removed in that regime. This does not prove that the conformal factor is unphysical, but it does suggest that absolute local scale may be less fundamental than causal structure at high curvature. Weyl invariance may therefore provide a natural organising principle for the ultraviolet theory.

There is also an older geometric motivation for this idea, already emphasised by Weyl~\cite{Weyl:1918ib}. In Riemannian geometry, the direction of a vector at separated points is compared only infinitesimally through parallel transport, whereas its magnitude is treated as directly comparable across spacetime once a metric is given. Weyl regarded this asymmetry as conceptually ad hoc. If vector lengths are only meaningfully comparable locally, rather than absolutely across spacetime, then the local normalization of the metric may be better understood as surplus structure, suggesting a Weyl-invariant description of geometry.

A complementary, more field-theoretic motivation comes from the renormalization group. In quantum field theory, a genuine ultraviolet completion controlled by a fixed point is expected to exhibit scale-invariant behaviour at short distances~\cite{Litim:2003vp,Shomer:2007vq,Weinberg:1976xy}. In gravity, this idea is especially suggestive because the metric itself carries the notion of local scale. In an ordinary non-gravitational quantum field theory, one may discuss the running of couplings relative to a fixed background geometry. In gravity, that interpretation is more subtle because the notion of distance is itself dynamical. If absolute local scale ceases to be fundamental in the ultraviolet, it is natural to ask whether gravity approaches a Weyl-invariant regime.

The Palatini formulation of $f(\mathcal{R})$ gravity provides a particularly simple setting in which to explore this possibility~\cite{Sotiriou:2008rp,Olmo:2011uz}. In this formulation, the metric and affine connection are treated as independent variables, and the independent connection is taken to be inert under Weyl rescalings of the metric. The Palatini scalar curvature then transforms homogeneously as $\RPal\longrightarrow\Omega^{-2}\mathcal{R}$, so that in four dimensions the density $\sqrt{-g}\mathcal{R}^{2}$ is exactly Weyl invariant~\cite{Edery:2019txq}. An additional attraction of the Palatini formulation is that it does not introduce the fourth-order metric equations characteristic of metric quadratic gravity and therefore generically avoids the corresponding higher-derivative spin-2 instability~\cite{Stelle:1977ry}. The Palatini $\mathcal{R}^{2}$ theory is consequently a natural candidate for an ultraviolet endpoint.

These observations motivate asymptotically Weyl-invariant gravity (AWIG)~\cite{Coumbe:2019fht,Coumbe:2021qid,Coumbe:2025ktl}\interfootnotelinepenalty=10000
\footnote{\scriptsize For earlier motivation see Refs.~\cite{Coumbe:2015bka,Coumbe:2018myj,Coumbe:2015aev,Coumbe:2015zqa}.}, defined by the action

\begin{equation}
S=\alpha\int d^4x\,\sqrt{-g}\,\mathcal{R}^{\,n(\mathcal{R}_{*})},
\qquad
n(\mathcal{R_{*}})\to
\begin{cases}
1, & \mathcal{R_{*}}\to 0^+,\\
2, & \mathcal{R_{*}}\to +\infty,
\end{cases}
\label{eq:AWIG_action_intro}
\end{equation}

\noindent where $\mathcal{R}$ is the Palatini scalar curvature, with the metric and connection treated as independent variables~\cite{Edery:2019txq}. The exponent $n(\mathcal{R}_{*})$ is dimensionless, as signified by the dimensionless argument $\mathcal{R}_{*}\equiv \mathcal{R}/\mathcal{R}_{0}$ used throughout, with $\mathcal{R}_{0}>0$ a fixed reference curvature of mass dimension two~\cite{Coumbe:2019fht}. The curvature term $\mathcal{R}^{\,n(\mathcal{R_{*}})}$ has canonical mass dimension $2n(\mathcal{R}_{*})$, so the effective inverse gravitational coupling $\alpha$ has canonical mass dimension $4-2n(\mathcal{R}_{*})$. At the infrared endpoint $n=1$, $\alpha$ therefore has canonical mass dimension two and the theory approaches the Einstein--Hilbert action~\cite{Sotiriou:2008rp}. At the ultraviolet endpoint $n=2$, $\alpha$ becomes dimensionless and AWIG approaches the Weyl-invariant \(\mathcal{R}^2\) theory~\cite{Edery:2019txq}. In this framework, the exponent \(n(\mathcal{R}_{*})\) plays the role of an effective running quantity. Rather than allowing Newton's constant to run relative to an externally fixed notion of scale~\cite{Anber:2011ut}, the scale dependence is encoded directly in the curvature-scaling law itself.\interfootnotelinepenalty=10000 \footnote{\scriptsize This should not, however, be interpreted as a microscopic renormalization-group flow unless such a relation is derived independently.}

AWIG is built on the hypothesis that Weyl invariance governs the ultraviolet regime of gravity. Far from being arbitrary, this hypothesis is highly restrictive: within the minimal scalar Palatini theory space defined below, it uniquely selects the \(\RPal^2\) ultraviolet endpoint rather than leaving the high-curvature functional form undetermined. The proposal also retains the Einstein--Hilbert form in the infrared and avoids introducing the higher-derivative spin-2 mode of generic metric quadratic gravity. Previous studies have examined the viability and falsifiability of AWIG in concrete physical settings~\cite{Coumbe:2021qid,Coumbe:2025ktl}; the present work addresses the complementary question of whether its symmetry-selected ultraviolet endpoint has a controlled counterterm structure. The physical value of the framework is therefore testable rather than assumed: its interpolation and matter coupling must ultimately yield predictions distinguishable from general relativity and from other modified-gravity theories.

The main purpose of this paper is to advance our understanding of the ultraviolet structure of AWIG, building on the work of Refs.~\cite{Coumbe:2015bka,Coumbe:2018myj,Coumbe:2015aev,Coumbe:2015zqa,Coumbe:2019fht,Coumbe:2021qid,Coumbe:2025ktl}. First, we show that within the minimal scalar Palatini $f(\RPal)$ sector, exact Weyl invariance uniquely selects the $\RPal^{2}$ action as the ultraviolet endpoint. Second, under explicitly phenomenological assumptions, we constrain a class of smooth autonomous curvature-flow functions compatible with the required infrared and ultraviolet endpoints. Finally, we analyse the strict ultraviolet limit $n(\RPal_{*})\to2$ and show that, under a controlled set of assumptions, the corresponding Palatini $\RPal^{2}$ theory exhibits conditional on-shell closure of the restricted ultraviolet counterterm sector at arbitrary loop order, modulo the Euler term, on the regular branch $\RPal\neq0$.

This does not establish full perturbative renormalizability. In particular, the result is restricted to a specified curvature-built theory space, is obtained on shell, and is conditional on the absence of anomalies obstructing the classical Weyl and projective symmetries. It nevertheless provides evidence for improved ultraviolet behaviour relative to Einstein gravity, beyond AWIG's already established power-counting renormalizability~\cite{Coumbe:2025ktl}. In Einstein gravity, by contrast, new non-renormalizable counterterm structures arise at higher loop order~\cite{tHooft:1974bx,Goroff:1985th}.

The distinction between off-shell and on-shell will be important in what follows. Off-shell, one studies the action and its symmetry properties for arbitrary field configurations, without imposing the equations of motion. On-shell, by contrast, one restricts to field configurations that satisfy the classical equations of motion~\cite{Brouder:2007db,Arkani-Hamed:2012zlh}.

\section{How AWIG in the ultraviolet may uniquely satisfy local scale invariance}

Although Eq.~(\ref{eq:AWIG_action_intro}) records the endpoint conditions defining AWIG, the following argument does not assume the UV value $n=2$. Instead, it begins with a general function $f(\RPal)$ in the minimal scalar Palatini sector and asks which form is independently selected by exact Weyl invariance.

In the Palatini formulation considered here, the metric \(g_{\mu\nu}\) and affine connection \(\Gamma^{\lambda}{}_{\mu\nu}\) are independent variables~\cite{Olmo:2011uz}. We take Weyl transformations to act only on the metric,
\begin{equation}
    g_{\mu\nu}(x)\to \tilde g_{\mu\nu}(x)=\Omega^2(x)g_{\mu\nu}(x),
    \qquad
    \Gamma^{\lambda}{}_{\mu\nu}\to \tilde\Gamma^{\lambda}{}_{\mu\nu}=\Gamma^{\lambda}{}_{\mu\nu},
    \qquad \Omega(x)>0.
    \label{eq:weyl-pal}
\end{equation}
Because the connection is held fixed, the Palatini Ricci tensor \(\Ric\) is unchanged, while the inverse metric transforms as \(g^{\mu\nu}\to\Omega^{-2}g^{\mu\nu}\). It follows that the Palatini scalar curvature
\begin{equation}
    \RPal(g,\Gamma):=g^{\mu\nu}\mathcal{R}_{\mu\nu}(\Gamma)
\end{equation}
transforms homogeneously,
\begin{equation}
    \RPal(g,\Gamma)\to \RPal(\tilde g,\Gamma)=\Omega^{-2}\RPal(g,\Gamma),
\end{equation}
whereas in four dimensions the volume density transforms as \(\sqrt{-g}\to\Omega^4\sqrt{-g}\). Therefore
\begin{equation}
    \sqrt{-g}\,\RPal^2\to \sqrt{-g}\,\RPal^2,
\end{equation}
so the Palatini \(\RPal^2\) density is exactly Weyl invariant in four dimensions~\cite{Edery:2019txq}.

This observation can be sharpened into a uniqueness statement, provided one specifies the theory space. Consider the minimal scalar Palatini sector, namely, actions of the form
\begin{equation}
    S[g,\Gamma]=\int d^4x\,\sqrt{-g}\,f(\RPal),
    \label{eq:min-sector}
\end{equation}
with no explicit dependence on derivatives of \(\RPal\), no explicit insertions of nonmetricity or torsion, and no additional curvature contractions. Under a Weyl transformation, one has
\begin{equation}
    \sqrt{-g}\,f(\RPal)\to \Omega^4\sqrt{-g}\,f(\Omega^{-2}\RPal).
\end{equation}

\noindent Exact Weyl invariance therefore requires~\cite{Edery:2019txq}
\begin{equation}
    \Omega^4 f(\Omega^{-2}\RPal)=f(\RPal),
    \qquad \text{for all $\RPal$ and all $\Omega>0$}.
\end{equation}
Defining \(\lambda:=\Omega^{-2}\), this becomes
\begin{equation}
    f(\lambda \RPal)=\lambda^{2}f(\RPal),
    \qquad \text{for all $\lambda>0$}.
\end{equation}
This is precisely the condition that \(f\) be homogeneous of degree 2. Assuming ordinary regularity, the unique solution is
\begin{equation}
    f(\RPal)=A \RPal^2,
\end{equation}
with $A$ constant. Thus, within the minimal scalar Palatini \(f(\RPal)\) sector, exact Weyl invariance uniquely selects the \(\RPal^2\) action in four dimensions.

This explains the ultraviolet structure of AWIG. Using the dimensional convention introduced in the Introduction, the endpoint conditions are
\begin{equation}
    n(\mathcal{R}_{*})\to1
    \quad (\mathcal{R}_{*}\to0^+),
    \qquad
    n(\mathcal{R}_{*})\to2
    \quad (\mathcal{R}_{*}\to+\infty).
\end{equation}
The complete interpolating function in Eq.~\eqref{eq:AWIG_action_intro} is not homogeneous of degree two when \(n\) varies with \(\mathcal{R}_{*}\). The homogeneity argument above applies only to the exact ultraviolet endpoint function, for which \(n=2\) is constant and
\begin{equation}
    f_{\mathrm{UV}}(\mathcal{R})=A\mathcal{R}^2.
\end{equation}
Within the minimal scalar Palatini sector, this is the only exactly Weyl-invariant ultraviolet endpoint. In that precise sense, the endpoint \(n(\mathcal{R}_{*})\to2\) is uniquely selected by the symmetry requirement off shell; the intermediate interpolation is not fixed by that argument.

Later, in Sec.~4, this uniqueness result will be partially strengthened on-shell. Under additional assumptions, the same \(\mathcal{R}^2\) structure re-emerges on shell even within a broader restricted ultraviolet sector that is not limited to \(f(\mathcal{R})\) models. Thus, within the exact setting defined in Sec.~4, \(\mathcal{R}^2\) also turns out to be the unique non-topological on-shell representative of the ultraviolet Weyl-invariant sector.

Taken together, these observations sharpen the ultraviolet interpretation of AWIG. Exact Weyl invariance may uniquely select \(\mathcal{R}^2\) as the ultraviolet endpoint action, so the limit \(n(\mathcal{R}_{*})\to2\) is not merely a convenient ansatz but the unique symmetry-selected endpoint in that theory space. If this picture is correct, then AWIG may be characterised as the interpolation between two symmetry-defined endpoints---general relativity in the infrared and Weyl invariance in the ultraviolet.

\section{$n(\mathcal{R}_{*})$ as an effective curvature-dependent quantity}

In ordinary quantum field theory, renormalization-group flow is typically formulated relative to a fixed background spacetime, or at least relative to a fixed external notion of scale. Gravity is different. The gravitational field is the spacetime geometry, and the metric \(g_{\mu\nu}\) itself determines distances, times, and causal structure. Once the metric becomes dynamical, the notion of scale is no longer simply external, but becomes tied to the geometry whose dynamics one is trying to describe.

This creates a conceptual tension in approaches that parametrise gravitational ultraviolet behaviour through a running Newton coupling \(G(k)\)~\cite{Anber:2011ut}. In many applications, the renormalization-group parameter \(k\) is interpreted through an effective inverse length or momentum scale, but in gravity that interpretation is more subtle because geometry is itself dynamical. Closely related concerns have been emphasised by Donoghue and collaborators~\cite{Anber:2011ut,Donoghue:2019clr}, who argue more narrowly that in perturbative low-energy gravity there is no unique, universal, process-independent notion of a running Newton constant directly analogous to the running couplings of renormalizable gauge theories. Different observables can receive different quantum corrections, so a single function \(G(k)\) need not provide an invariant description across processes~\cite{Anber:2011ut,Donoghue:2019clr}.

AWIG shifts the locus of scale dependence from the coupling strength to the curvature-scaling law itself. In the ultraviolet, where \(n\to2\), the theory approaches the Weyl-invariant endpoint; in the infrared, where \(n\to1\), it recovers the Einstein-like action~\cite{Sotiriou:2008rp}. However, the dimensionless local scalar \(\mathcal{R}_{*}\) is not a Wilsonian coarse-graining scale. The analysis below is therefore phenomenological: it asks what follows if the curvature dependence of \(n\) can be represented by a smooth autonomous first-order equation. It does not derive the quantum renormalization-group flow of the theory.

\subsection{A class of smooth effective flow functions}

Define
\begin{equation}
    \beta_{\mathrm{eff}}(n):=
    \frac{dn}{d\ln \mathcal{R}_{*}}.
\end{equation}
Here \(\beta_{\mathrm{eff}}\) is an effective curvature-flow function, not a beta function obtained by integrating quantum fluctuations. If one assumes that the interpolation is autonomous and analytic in a neighbourhood of the interval connecting \(n=1\) and \(n=2\), and that these endpoint values are zeros of the flow, analytic factorisation gives
\begin{equation}\label{beta1}
    \beta_{\mathrm{eff}}(n)
    =(n-2)^a(n-1)^b p(n),
\end{equation}
where \(a,b\in\mathbb N\) are the multiplicities of the two zeros and \(p(n)\) is analytic and nonvanishing over \([1,2]\). Equation~\eqref{beta1} is therefore the most general local analytic form under the stated assumptions. Neither autonomy nor analyticity is established by the underlying quantum theory.

\subsection{The ultraviolet endpoint}

To characterise the approach to the ultraviolet Weyl-invariant endpoint, it is useful to consider the combination appearing in the Palatini trace equation~\cite{Sotiriou:2008rp}
\begin{equation}\label{traceEq}
    T(\mathcal{R})
    :=f'(\mathcal{R})\,\mathcal{R}-2f(\mathcal{R}).
\end{equation}
At the exact ultraviolet endpoint, Section~2 gives
$f_{\rm UV}(\mathcal R)=A\mathcal R^2$, which is differentiable and
homogeneous of degree two. Euler's theorem therefore implies
\begin{equation}
    \mathcal{R}\,f_{\mathrm{UV}}'(\mathcal{R})
    =2f_{\mathrm{UV}}(\mathcal{R}).
\end{equation}
Thus \(T=0\) identically for the exact Palatini \(\mathcal{R}^2\) theory. The full AWIG interpolation is not homogeneous, so \(T\) is generally nonzero away from the endpoint. For the interpolating function in Eq.~\eqref{eq:AWIG_action_intro}, with logarithms of curvature understood in terms of the dimensionless variable \(\RPal_*\), one finds
\begin{equation}
T(\RPal)
=
\RPal^{\,n(\RPal_*)}
\left[
(n-2)
+
(\ln\RPal_*)\beta_{\mathrm{eff}}(n)
\right].
\label{tofr}
\end{equation}
Requiring \(T\to0\) as \(\mathcal{R}_{*}\to+\infty\) expresses the stronger condition that the full function, rather than only the exponent, approaches the homogeneous \(\mathcal{R}^2\) endpoint.\\

\noindent\textbf{Proposition 3.1.}\\
Let
\begin{equation}
\frac{dn}{d\ln \mathcal{R}_{*}}
=\beta_{\mathrm{eff}}(n)
:=(n-2)^a(n-1)^b p(n),
\end{equation}
where \(a,b\in\mathbb N\), \(p(n)\) is analytic in a neighbourhood of \(n=2\), and \(n(\mathcal{R}_{*})\) satisfies
\begin{equation}
\lim_{\mathcal{R}_{*}\to0^+}n(\mathcal{R}_{*})=1,
\qquad
\lim_{\mathcal{R}_{*}\to+\infty}n(\mathcal{R}_{*})=2.
\end{equation}
If \(T(\mathcal{R})\to0\) as \(\mathcal{R}_{*}\to+\infty\), then necessarily
\begin{equation}
    a=1,
    \qquad
    p(2)<-2.
\end{equation}

\noindent\textbf{Proof of Proposition 3.1.}\\
Set
\begin{equation}
    t:=\ln\mathcal{R}_{*},
    \qquad
    \epsilon(t):=2-n(t),
\end{equation}
so that \(\epsilon(t)\to0^+\) as \(t\to+\infty\). Since \(p(n)\) is analytic at \(n=2\),
\begin{equation}
    \beta_{\mathrm{eff}}(n)
    =(-1)^a p(2)\epsilon^a
    +O(\epsilon^{a+1}).
\end{equation}
Because \(\epsilon=2-n\),
\begin{equation}
    \frac{d\epsilon}{dt}
    =-\beta_{\mathrm{eff}}(n)
    =-(-1)^a p(2)\epsilon^a
    +O(\epsilon^{a+1}).
\end{equation}
If \(a>1\), then to leading order
\begin{equation}
    \frac{d\epsilon}{dt}\sim K\epsilon^a,
\end{equation}
with \(K\neq0\). Separating variables and integrating gives \(\epsilon(t)\sim t^{-1/(a-1)}\). The bracket in Eq.~\eqref{tofr} then decays only as a power of \(t\), whereas, since \(\RPal=\RPal_0 e^t\),
\begin{equation}
\RPal^{\,n(t)}
=
\RPal_0^{\,n(t)}e^{n(t)t}
\sim
\RPal_0^2 e^{2t}
\qquad
(t\to+\infty).
\label{eq:UV_growth}
\end{equation}
Therefore \(T(\RPal)\) cannot tend to zero, and \(a>1\) is
excluded.

For \(a=1\),
\begin{equation}
    \frac{d\epsilon}{dt}
    =p(2)\epsilon+O(\epsilon^2),
\end{equation}
so
\begin{equation}
    \epsilon(t)=Ce^{p(2)t}(1+o(1)).
\end{equation}
Convergence requires \(p(2)<0\). Substitution into Eq.~\eqref{tofr} gives
\begin{equation}
    T(\mathcal{R})
    =\mathcal{R}_0^2 C e^{(2+p(2))t}
    \bigl[-1-p(2)t\bigr](1+o(1)).
\end{equation}
This tends to zero only if \(2+p(2)<0\), and hence \(p(2)<-2\).

\subsection{The infrared endpoint}

To motivate the infrared counterpart, we now impose a minimal nondegeneracy condition on the approach to the Einstein-like infrared endpoint at $n=1$. If $n(\mathcal{R}_{*})$ is continuously differentiable ($C^1$) in a neighbourhood of $\mathcal{R}_{*}=0$ and satisfies
\begin{equation}
n'(0)=C\neq 0,
\end{equation}
then the leading departure from the infrared endpoint is linear,
\begin{equation}
n(\mathcal{R}_{*}) = 1 + C\mathcal{R}_{*} + o(\mathcal{R}_{*})
\qquad (\mathcal{R}_{*}\to 0).
\end{equation}
This is the simplest smooth way in which the exponent can move away from $n=1$ while still recovering Einstein scaling in the infrared. It excludes more degenerate possibilities in which the flow approaches the endpoint more flatly, and therefore provides a natural criterion for isolating the case of a simple zero of the effective infrared flow.\\

\noindent\textbf{Proposition 3.2.}\\
Let \(n(\mathcal{R}_{*})\in C^1([0,\delta))\) satisfy
\begin{equation}
\lim_{\mathcal{R}_{*}\to0^+}n(\mathcal{R}_{*})=1,
\qquad
\left.\frac{dn}{d\mathcal{R}_{*}}\right|_{0}=C\neq0.
\end{equation}
Assume that for all sufficiently small \(\mathcal{R}_{*}>0\),
\begin{equation}
\frac{dn}{d\ln \mathcal{R}_{*}}
=(n-2)^a(n-1)^b p(n),
\end{equation}
where \(p(n)\) is continuous at \(n=1\), \(p(1)\neq0\), and \(a=1\). Then necessarily
\begin{equation}
    b=1,
    \qquad p(1)=-1.
\end{equation}

\noindent\textbf{Proof of Proposition 3.2.}\\
The assumed expansion gives
\begin{equation}
    n(\mathcal{R}_{*})
    =1+C\mathcal{R}_{*}+o(\mathcal{R}_{*}),
\end{equation}
and therefore
\begin{equation}
    \frac{dn}{d\ln\mathcal{R}_{*}}
    =\mathcal{R}_{*}\frac{dn}{d\mathcal{R}_{*}}
    =C\mathcal{R}_{*}+o(\mathcal{R}_{*}).
\end{equation}
Hence
\begin{equation}
    \lim_{\mathcal{R}_{*}\to0^+}
    \frac{\frac{dn}{d\ln\mathcal{R}_{*}}}
    {n(\mathcal{R}_{*})-1}=1.
\end{equation}
On the other hand,
\begin{equation}
\frac{\frac{dn}{d\ln\mathcal{R}_{*}}}
{n(\mathcal{R}_{*})-1}
=(n-2)^a(n-1)^{b-1}p(n).
\end{equation}
Taking the infrared limit and using \(a=1\) gives
\begin{equation}\label{product}
1=-\lim_{\mathcal{R}_{*}\to0^+}
(n(\mathcal{R}_{*})-1)^{b-1}
 p(n(\mathcal{R}_{*})).
\end{equation}
Because \(p(1)\neq0\), the limit is finite and nonzero only when \(b=1\). It then follows that \(p(1)=-1\).

\subsection{A representative for $n(\mathcal{R}_{*})$}

A simple smooth family reproducing the endpoint behaviour motivated by Propositions~3.1 and~3.2, namely that $a=b=1$, $p(2)<-2$ and $p(1)=-1$, is
\begin{equation}
n(\RPal_*)=2-\frac{1}{1+c_1\RPal_*+c_3\RPal_*^3},
\qquad c_1>0,\quad c_3>0.
\end{equation}

This family may be motivated by two elementary exactly integrable model flows. Since $p(1)=-1$, if one takes

\begin{equation}
\beta_{\mathrm{eff}}(n)=-(n-2)(n-1),
\end{equation}

\noindent then direct integration gives

\begin{equation}
n(\RPal_*)=2-\frac{1}{1+C\RPal_*},
\end{equation}

\noindent which has the desired simple infrared behaviour
$n(\RPal_*)=1+C\RPal_*+O(\RPal_*^2)$ but only a relatively mild ultraviolet falloff. Note that this functional form was also independently derived via dynamical dimensional reduction of the spectral dimension in Ref.~\cite{Coumbe:2025ktl}, modulo a derivative term.

By contrast, since $p(2)<-2$, if one takes

\begin{equation}
\beta_{\mathrm{eff}}(n)=-3(n-2)(n-1),
\end{equation}

\noindent then integration yields

\begin{equation}
n(\RPal_*)=2-\frac{1}{1+C\RPal_*^3},
\end{equation}

\noindent which has a stronger ultraviolet approach to $n=2$ but a degenerate infrared limit with

\begin{equation}
\left.\frac{dn}{d\RPal_*}\right|_{\RPal_*=0}=0.
\end{equation}

The family

\begin{equation}
n(\RPal_*)=2-\frac{1}{1+c_1\RPal_*+c_3\RPal_*^3},
\qquad c_1>0,\quad c_3>0,
\end{equation}

\noindent may therefore be viewed as a simple smooth interpolant combining the nondegenerate infrared behaviour with the stronger ultraviolet falloff. The endpoint analysis does not fix the coefficients $c_1$ and $c_3$. The choice $c_1=c_3=1$ gives the particularly simple representative

\begin{equation}
  n(\RPal_*)=2-\frac{1}{1+\RPal_* +\RPal_*^3}.
\end{equation}

\subsection{Relation to a functional-renormalization-group calculation}

A first-principles derivation would require a quantum effective action with an independent metric and connection. One possible route is to introduce an effective average action \(\Gamma_k[g,\Gamma]\)~\cite{Reuter:1996cp}, expand both fields about suitable backgrounds, fix the diffeomorphism and projective gauge freedoms, include the corresponding ghost operators, and project the resulting functional flow onto an appropriate Palatini curvature truncation~\cite{Gies:2022ikv}. Such a calculation would determine whether the endpoint structure assumed above is realised dynamically and whether the exponent can be consistently interpreted as an effective coordinate on the quantum theory space. It could also generate derivative, torsion, nonmetricity, or additional curvature operators not contained in the present ansatz~\cite{Melichev:2024MAG}. Such a lengthy calculation is beyond the scope of the present work, but remains an important future step. The curvature-flow analysis should therefore be regarded as a constrained phenomenological parametrisation whose validity remains to be tested.

\section{Ultraviolet behaviour of AWIG}

\subsection{Setup and restricted counterterm basis}

We consider the strict ultraviolet theory~\cite{Edery:2019txq}
\begin{equation}
S[g,\Gamma]=\alpha\int d^4x\,\sqrt{-g}\,\RPal(g,\Gamma)^2,
\qquad
\RPal(g,\Gamma):=g^{\mu\nu}\RPal_{\mu\nu}(\Gamma),
\end{equation}
where $\Gamma^\lambda{}_{\mu\nu}$ is a torsionless affine connection independent of $g_{\mu\nu}$.

We work within a restricted ultraviolet theory space that is formalised precisely in  Proposition~4.1 below. In brief, we consider local, polynomial, parity-even scalar densities of canonical dimension four, built algebraically from the metric $g_{\mu\nu}$ and the affine curvature $\mathcal{R}^\lambda{}_{\rho\mu\nu}(\Gamma)$, with no explicit insertions of torsion, nonmetricity, derivatives of curvature, or Levi-Civita tensor contractions. Admissible counterterms are required to be invariant under diffeomorphisms and Weyl rescalings, and to depend only on the projective class of $\Gamma$; we work throughout on a torsionless representative of that class. The ultraviolet endpoint action is the pure Palatini $\mathcal{R}^2$ theory, and the on-shell reduction is performed on the regular branch $\mathcal{R} \neq 0$.

A technical comment is in order concerning the relation between projective invariance and the torsionless restriction. In the Palatini \(f(\mathcal{R})\) sector, projective invariance is naturally defined on the full space of affine connections~\cite{Sotiriou:2006qn},
\begin{equation}
\Gamma^{\lambda}{}_{\mu\nu} \;\mapsto\; \Gamma^{\lambda}{}_{\mu\nu} + \delta^{\lambda}{}_{\mu}\,\xi_{\nu}.
\end{equation}
A generic projective transformation does not preserve torsionlessness~\cite{Bejarano:2019zco}. Accordingly, the logical order adopted below is the following. First, admissible counterterms are classified on the full space of affine connections subject to the stated restrictions of locality, polynomiality, canonical dimension four, parity-evenness, diffeomorphism invariance, Weyl invariance, and insensitivity to the projective direction. Only after this classification is performed do we choose a torsionless representative of each projective equivalence class in order to write the resulting invariants in a convenient form. Thus torsionlessness is used below only as a gauge choice inside a projective class, not as a subspace on which projective transformations are assumed to act~\cite{Bejarano:2019zco}. In particular, whenever projective invariance is invoked to eliminate dependence on the antisymmetric Ricci sector, the argument is understood to take place on the full projective orbit before passing to a torsionless representative.

The counterterm analysis below is carried out in a restricted ultraviolet theory space chosen to match the minimal field content and symmetry structure of the strict $n(\mathcal{R}_{*})\to 2$ theory. This restriction is not arbitrary. In four dimensions, the ultraviolet limit is the pure Palatini $\mathcal{R}^2$ theory. More precisely, the underlying Palatini curvature sector is diffeomorphism, projective, and Weyl invariant, while the counterterm analysis below is carried out on a torsionless representative of the corresponding projective class. A natural first test of its ultraviolet consistency is therefore to classify local polynomial scalar densities of canonical dimension four built only from $g_{\mu\nu}$ and $\RPal^\lambda{}_{\rho\mu\nu}(\Gamma)$, without enlarging the field content or introducing tensorial structures absent from the ultraviolet action itself.

Excluding explicit torsion, nonmetricity insertions, derivatives of curvature, and parity-odd Levi-Civita contractions is likewise not meant to claim that such terms can never arise in a more general metric-affine completion. Rather, it isolates the minimal curvature-built sector continuously connected to the pure Palatini $\RPal^2$ endpoint, and therefore the sector in which the closure question can be posed most sharply. The result obtained below should be read in this precise sense, not as a classification of all possible metric-affine counterterms, but as a test of whether the ultraviolet Weyl-invariant core of the theory is already self-consistent on shell within its own minimal sector.

A further question concerns how quantisation may affect the classical symmetry structure. A natural assumption is that, in the absence of anomalies, the local symmetries of the classical ultraviolet action continue to constrain the divergent part of the quantum effective action. This follows the standard effective-field-theory expectation: ultraviolet divergences are local, and when quantisation preserves the exact gauge symmetries of the classical action, the allowed counterterms are organised by those same symmetries~\cite{Barvinsky:2017zlx}. Under this minimal assumption, the endpoint interpretation has content. The argument below is therefore conditional; it classifies divergences under the assumption that no anomaly in the measure, gauge-fixing procedure, ghost sector, or regularisation obstructs these symmetries~\cite{Barvinsky:2017zlx}.\\

\noindent \textbf{Proposition 4.1.}\\
Let $M$ be a four-dimensional manifold equipped with a nondegenerate metric $g_{\mu\nu}$ and an affine connection $\Gamma^\lambda{}_{\mu\nu}$. Assume throughout that:
\begin{enumerate}
\item one works on a projective equivalence class of affine connections which admits a torsionless representative;
\item admissible local counterterm densities are scalar densities of the form $\sqrt{-g}\,\Phi$, where $\Phi$ is a local polynomial scalar satisfying all of the following conditions:
\begin{enumerate}
\item in any chosen torsionless representative of the projective class, $\Phi$ is represented as an algebraic polynomial in $g_{\mu\nu}$ and in the curvature tensor $\mathcal{R}^\lambda{}_{\rho\mu\nu}(\Gamma)$;
\item $\Phi$ contains no derivatives of curvature, no explicit torsion insertions, no explicit nonmetricity insertions, and no Levi--Civita tensor insertions;
\item $\Phi$ is parity-even;
\item $\Phi$ has canonical mass dimension four;
\item $\sqrt{-g}\,\Phi$ is invariant under diffeomorphisms and under Weyl rescalings
\[
g_{\mu\nu}\mapsto \Omega^2 g_{\mu\nu},\qquad
\Gamma^\lambda{}_{\mu\nu}\mapsto \Gamma^\lambda{}_{\mu\nu},
\qquad \Omega>0;
\]
\item $\Phi$ is projectively invariant, in the sense that it depends only on the projective class of $\Gamma$;
\end{enumerate}
\item the on-shell reduction is performed in the pure Palatini $\mathcal{R}^2$ theory
\begin{equation}
S[g,\Gamma]=\alpha\int d^4x\,\sqrt{-g}\,\mathcal{R}^2
\end{equation}
on the regular branch $\mathcal{R}\neq 0$.
\end{enumerate}
Define
\begin{equation}
\mathcal{R}_{\mu\nu}:=\mathcal{R}^\lambda{}_{\mu\lambda\nu},\qquad
\mathcal{R}:=g^{\mu\nu}\mathcal{R}_{\mu\nu},\qquad
S_{\mu\nu}:=\mathcal{R}_{(\mu\nu)}-\frac14 g_{\mu\nu}\mathcal{R}.
\end{equation}
Then, on each connected component of the regular branch, every admissible counterterm density reduces on shell to a linear combination of $\sqrt{-g}\,\mathcal{R}^2$ and $\sqrt{-h}\,E_4(h)$, where $E_4(h)$ is the Euler density of the metric $h_{\mu\nu}$, and $h_{\mu\nu}:=|\mathcal{R}|\,g_{\mu\nu}$. Equivalently, for every admissible $\Phi$ there exist constants $A$ and $B$ such that on shell
\begin{equation}
\sqrt{-g}\,\Phi \;\approx\; A\,\sqrt{-g}\,\mathcal{R}^2 + B\,\sqrt{-g}\,E_4(g).
\end{equation}

\noindent \textbf{Proof of Proposition 4.1.}\\
The proof is local. Fix a connected open set $U\subset M$ contained in the regular branch $\mathcal{R}\neq 0$. Since $\mathcal{R}$ is continuous and nowhere vanishing on $U$, its sign is constant there. Write
\begin{equation}
\sigma:=\operatorname{sgn}(\mathcal{R})\in\{+1,-1\},
\qquad
h_{\mu\nu}:=|\mathcal{R}|\,g_{\mu\nu}=\sigma \mathcal{R}\,g_{\mu\nu}.
\end{equation}

\noindent \emph{Step 1: Canonical dimension forces quadratic curvature dependence.}\\
Use the standard effective-field-theory assignments
\begin{equation}
[x^\mu]=-1,\qquad [\partial_\mu]=+1,\qquad [g_{\mu\nu}]=0.
\end{equation}
Since
\begin{equation}
[\mathcal{R}^\lambda{}_{\rho\mu\nu}]=2,
\end{equation}
and $\Phi$ is algebraic in the curvature, contains no derivatives of curvature, and has canonical mass dimension four (by assumptions~2(a), (b) and (d) of Proposition~4.1), it follows that $\Phi$ is homogeneous of degree two in the curvature. Hence every admissible counterterm density is quadratic in $\mathcal{R}^\lambda{}_{\rho\mu\nu}(\Gamma)$.\\

\noindent \emph{Step 2: Metric field equation on the regular branch.}\\
Variation of
\[
S[g,\Gamma]=\alpha\int d^4x\,\sqrt{-g}\,\mathcal{R}^2
\]
with respect to $g_{\mu\nu}$ gives
\begin{equation}
2\mathcal{R}\,\mathcal{R}_{(\mu\nu)}-\frac12 g_{\mu\nu}\mathcal{R}^2=0.
\end{equation}
Equivalently,
\begin{equation}
2\mathcal{R}\,S_{\mu\nu}=0.
\end{equation}
On the regular branch $\mathcal{R}\neq 0$ (assumption~3 of Proposition~4.1), this implies
\begin{equation}
S_{\mu\nu}=0,
\qquad
\mathcal{R}_{(\mu\nu)}=\frac14 g_{\mu\nu}\mathcal{R}.
\end{equation}
Using $h_{\mu\nu}=\sigma \mathcal{R}\,g_{\mu\nu}$, and because $\sigma^{-1}=\sigma$ this becomes
\begin{equation}
\mathcal{R}_{(\mu\nu)}=\frac{\sigma}{4}\,h_{\mu\nu}.
\end{equation}

\noindent \emph{Step 3: Connection field equation and projective gauge fixing.}\\
The Palatini $\mathcal{R}^2$ action is projectively invariant, so the connection is determined by its
equation of motion only up to projective transformations~\cite{Sotiriou:2006qn}. The Euler--Lagrange equation obtained
by varying the action with respect to the connection implies, up to this standard projective
indeterminacy, the compatibility condition~\cite{Olmo:2011uz}
\begin{equation}
\nabla^\Gamma_\lambda\!\left(\sqrt{-g}\,\mathcal{R}\,g^{\mu\nu}\right)=0.
\end{equation}

Since $\mathcal{R}$ is continuous and nowhere vanishing on $U$, its sign
\begin{equation}
\sigma:=\operatorname{sgn}(\mathcal{R})\in\{+1,-1\}
\end{equation}
is constant there. Define
\begin{equation}
h_{\mu\nu}:=|\mathcal{R}|\,g_{\mu\nu}.
\end{equation}
In four dimensions one has
\begin{equation}
\sqrt{-h}\,h^{\mu\nu}=\sqrt{-g}\,|\mathcal{R}|\,g^{\mu\nu},
\end{equation}
and since $\mathcal{R}=\sigma |\mathcal{R}|$ with constant $\sigma$ on $U$, the above compatibility condition is
equivalently
\begin{equation}
\nabla^\Gamma_\lambda\!\left(\sqrt{-h}\,h^{\mu\nu}\right)=0,
\end{equation}
again up to projective freedom.

By Assumption 1 of Proposition~4.1, the projective class of $\Gamma$ admits a torsionless representative. Choose
such a representative. For a torsionless connection, the condition
\[
\nabla^\Gamma_\lambda\!\left(\sqrt{-h}\,h^{\mu\nu}\right)=0
\]
is equivalent to
\begin{equation}
\nabla^\Gamma_\lambda h_{\mu\nu}=0,
\end{equation}
and therefore uniquely fixes that representative to be the Levi--Civita connection of $h$:
\begin{equation}
\Gamma^\lambda_{\mu\nu}=\Gamma^\lambda_{\mu\nu}(h).
\end{equation}
Hence, on shell and after fixing the projective gauge to a torsionless representative, the
independent connection reduces to the Levi--Civita connection of the conformally related metric
$h_{\mu\nu}=|\mathcal{R}|g_{\mu\nu}$~\cite{Edery:2019txq}.\\

\noindent \emph{Step 4: $h$ is Einstein with constant scalar curvature.}\\
Since $\Gamma=\Gamma(h)$ on shell, one has
\begin{equation}
\mathcal{R}_{\mu\nu}(\Gamma)=R_{\mu\nu}(h).
\end{equation}
Combining this with Step 2 yields
\begin{equation}
R_{\mu\nu}(h)=\frac{\sigma}{4}\,h_{\mu\nu}.
\end{equation}
Thus $h$ is Einstein. Taking the trace gives
\begin{equation}
R(h)=\sigma.
\end{equation}
In particular, $R(h)$ is constant and
\begin{equation}
R_{\mu\nu}(h)R^{\mu\nu}(h)=\frac14,
\qquad
R(h)^2=1.
\end{equation}

\noindent \emph{Step 5: Use projective invariance only after the on-shell reduction.}\\
Let $\sqrt{-g}\,\Phi$ be any admissible counterterm density. By assumption, $\Phi$ depends only on the projective class of $\Gamma$. Therefore, on shell, it may be evaluated on the torsionless Levi--Civita representative chosen in Step 3:
\begin{equation}
\sqrt{-g}\,\Phi(g,\Gamma)\;\approx\;\sqrt{-g}\,\Phi(g,\Gamma(h)).
\end{equation}
No prior classification of projectively invariant affine-curvature scalars is needed.\\

\noindent\emph{Step 6: Weyl invariance transfers the density from $g$ to $h$.}\\
Because $h_{\mu\nu}=|\mathcal{R}|\,g_{\mu\nu}$ and $|\mathcal{R}|>0$ on $U$, the metrics $g$ and $h$ are related by a genuine Weyl rescaling with factor $\Omega^2=|\mathcal{R}|$. Since $\sqrt{-g}\,\Phi$ is Weyl invariant for every fixed affine connection (assumption~2(e)), applying this Weyl rescaling with $\Gamma=\Gamma(h)$ held fixed gives
\begin{equation}
\sqrt{-g}\,\Phi(g,\Gamma(h))
=
\sqrt{-h}\,\Phi(h,\Gamma(h)).
\end{equation}
Therefore
\begin{equation}
\sqrt{-g}\,\Phi(g,\Gamma)\;\approx\;\sqrt{-h}\,\Phi(h,\Gamma(h)).
\end{equation}

\noindent \emph{Step 7: Classification in the Levi--Civita geometry of $h$.}\\
The right-hand side is now an ordinary parity-even local scalar density of canonical dimension four built algebraically from the metric $h_{\mu\nu}$ and its Levi--Civita curvature, with no derivatives of curvature and no Levi--Civita tensor insertions (by assumptions~2(a), 2(b), 2(c), and 2(d) of Proposition~4.1). In four dimensions, such quadratic scalar densities are linear combinations of~\cite{Fulling:1992vm}
\begin{equation}
\sqrt{-h}\,R_{\mu\nu\rho\sigma}(h)R^{\mu\nu\rho\sigma}(h),\qquad
\sqrt{-h}\,R_{\mu\nu}(h)R^{\mu\nu}(h),\qquad
\sqrt{-h}\,R(h)^2.
\end{equation}
Hence there exist constants $a,b,c$ such that on shell
\begin{equation}
\sqrt{-g}\,\Phi \;\approx\;
a\,\sqrt{-h}\,R_{\mu\nu\rho\sigma}(h)R^{\mu\nu\rho\sigma}(h)
+b\,\sqrt{-h}\,R_{\mu\nu}(h)R^{\mu\nu}(h)
+c\,\sqrt{-h}\,R(h)^2.
\end{equation}
Using Step 4,
\begin{equation}
R_{\mu\nu}(h)R^{\mu\nu}(h)=\frac14,
\qquad
R(h)^2=1,
\end{equation}
so this reduces to
\begin{equation}
\sqrt{-g}\,\Phi \;\approx\;
a\,\sqrt{-h}\,R_{\mu\nu\rho\sigma}(h)R^{\mu\nu\rho\sigma}(h)
+d\,\sqrt{-h}
\end{equation}
for some constant $d$.\\

\noindent \emph{Step 8: Replace the Riemann-squared term by the Euler density.}\\
The four-dimensional Euler density is~\cite{Lovelock:1971yv}
\begin{equation}
E_4(h)=
R_{\mu\nu\rho\sigma}(h)R^{\mu\nu\rho\sigma}(h)
-4R_{\mu\nu}(h)R^{\mu\nu}(h)
+R(h)^2.
\end{equation}
Using again Step 4,
\begin{equation}\label{canc}
-4R_{\mu\nu}(h)R^{\mu\nu}(h)+R(h)^2=-4\cdot\tfrac{1}{4} + 1=0,
\end{equation}
hence on shell
\begin{equation}
E_4(h)=R_{\mu\nu\rho\sigma}(h)R^{\mu\nu\rho\sigma}(h).
\end{equation}
Therefore
\begin{equation}
\sqrt{-g}\,\Phi \;\approx\; A\,\sqrt{-h} + B\,\sqrt{-h}\,E_4(h)
\end{equation}
for some (redefined) constants $A,B$.

The cancellation in Eq.~(\ref{canc}) is quite remarkable and deserves emphasis. This vanishing is not a numerical accident: it follows directly from the Einstein condition $\mathcal{R}_{\mu\nu}(h) = (\sigma/4)h_{\mu\nu}$, which is itself a consequence of the metric field equation of the Palatini $\mathcal{R}^2$ action on the regular branch $\mathcal{R} \neq 0$. The  identification of the Riemann-squared term with the Euler density is therefore a property specific to the $\mathcal{R}^2$ endpoint. Had the ultraviolet endpoint action been $\sqrt{-g}\,\mathcal{R}^p$ for $p \neq 2$, Weyl invariance would be lost and the density transfer of Step~6 would fail before this cancellation is even reached. The collapse of the counterterm sector is thus not a generic feature of Palatini theories, but a consequence of the same Weyl-invariance condition that uniquely selects $\mathcal{R}^2$ in Section~2.\\

\noindent \emph{Step 9: Rewrite $\sqrt{-h}$ as $\sqrt{-g}\,\mathcal{R}^2$.}\\
Since $h_{\mu\nu}=|\mathcal{R}|\,g_{\mu\nu}$ in four dimensions,
\begin{equation}
\sqrt{-h}=|\mathcal{R}|^2\sqrt{-g}=\mathcal{R}^2\sqrt{-g}.
\end{equation}
Since $E_4$ is topological, conformal rescalings change $\sqrt{-h}\,E_4(h)$ only by a boundary term, so $\sqrt{-h}\,E_4(h)$ and $\sqrt{-g}\,E_4(g)$ are equivalent as contributions to the effective action. Thus
\begin{equation}
\sqrt{-g}\,\Phi \;\approx\; A\,\sqrt{-g}\,\mathcal{R}^2 + B\,\sqrt{-g}\,E_4(g).
\end{equation}
This is the claimed restricted on-shell closure.

\subsection{Conditional on-shell closure}

Let $\Gamma_{\mathrm{div}}$ denote the divergent part of the effective action in the restricted sector under consideration. By Proposition~4.1, every admissible local divergence collapses on shell, on the regular branch $\mathcal{R} \neq 0$, to a linear combination of the original $\sqrt{-g}\,\mathcal{R}^2$ density and an Euler-density contribution constructed from the on-shell metric $h_{\mu\nu}=|\mathcal{R}|\,g_{\mu\nu}$. Accordingly,
\begin{equation}
\Gamma_{\mathrm{div}} \approx
c_1 \int d^4x\,\sqrt{-g}\,\mathcal{R}^2
+
c_2 \int d^4x\,\sqrt{-h}\,E_4(h),
\end{equation}
where $\approx$ denotes equality modulo the classical equations of motion, and $c_1,c_2$ are coefficients.

Thus, within the restricted torsionless, parity-even, curvature-built sector defined in Sec.~4.1, the ultraviolet theory is on-shell closed modulo the Euler-density contribution: all admissible divergences collapse to the original $\sqrt{-g}\,\mathcal{R}^2$ structure together with $\sqrt{-h}\,E_4(h)$.

The scope of this conclusion is limited but clear. First, it applies only on the regular branch $\mathcal{R}\neq 0$, where the auxiliary metric $h_{\mu\nu}=|\mathcal{R}|g_{\mu\nu}$ is well defined. Second, it is complete only within the restricted theory space of Sec.~4.1, and does not exclude additional dimension-four invariants in more general metric-affine extensions involving explicit torsion, nonmetricity, parity-odd terms, or derivatives of curvature. Third, the result is on shell: it does not by itself show that the divergent effective action is off-shell equivalent to a renormalization by local counterterms written purely in the original variables $(g_{\mu\nu},\Gamma^\lambda{}_{\mu\nu})$. Fourth, possible quantum anomalies obstructing the classical Weyl or projective symmetries are not addressed here.

We also make a note on the scope of the loop-order claim. The argument in Proposition~4.1 is an algebraic classification of admissible local invariants; it
does not depend on the loop order at which a divergence arises, because ultraviolet divergences are local at every order and must therefore lie in the space of local invariants consistent with the symmetries of the theory~\cite{Barvinsky:2017zlx}. The conclusion therefore applies at arbitrary loop order, subject to a single non-trivial assumption: that the classical Weyl and projective symmetries are not anomalous in the quantum theory.

This assumption is non-trivial. Weyl anomalies generically arise in quantum field theories coupled to curved spacetime~\cite{Duff:1993wm,Deser:1993yx}, and their presence or absence in the Palatini $R^2$ theory has not been established here. Concretely, one would need to verify that the path-integral measure, the gauge-fixing procedure for diffeomorphisms and projective transformations, and the resulting ghost sector preserve Weyl invariance at the quantum level, or that any Weyl variation of the measure is cancelled by a consistent anomaly-cancellation mechanism. If a Weyl anomaly is present, additional dimension-four invariants---not of the form $\sqrt{-g}R^2$ or $\sqrt{-h}E_4(h)$---could in principle appear as counterterms, and the on-shell closure result would not extend to the anomalous sector. The result established here is therefore conditional on anomaly absence, and the investigation of whether this condition holds is an important open problem.

Nevertheless, within the stated assumptions, the conclusion is nontrivial: the ultraviolet $n(\mathcal{R}_{*}) \to 2$ limit of AWIG exhibits conditional on-shell closure at arbitrary loop order in the restricted sector considered here. This provides genuine evidence for improved ultraviolet behaviour relative to Einstein gravity, while stopping short of establishing full off-shell perturbative renormalizability.

\section{Discussion and conclusions}

This paper aims to sharpen the theoretical formulation of AWIG as a candidate ultraviolet framework for gravity~\cite{Coumbe:2019fht,Coumbe:2021qid,Coumbe:2025ktl}. In particular, we clarify why Weyl invariance is a natural organising principle in the high-curvature regime, show how it selects the Palatini \(\RPal^2\) theory as the ultraviolet endpoint within the minimal scalar sector, and derive conditional asymptotic constraints on a phenomenological flow for \(n(\mathcal{R}_{*})\) between the infrared and ultraviolet endpoints. Within that sharpened framework, we then analyse the ultraviolet counterterm structure and show that, under explicit restrictions, the strict \(n(\mathcal{R}_{*})\to2\) limit exhibits conditional on-shell closure modulo the Euler term. Within this derivation, a remarkable cancellation reduces the Riemann-squared contribution to the Euler term, a feature unique to the \(\mathcal{R}^2\) endpoint.

In the framework of Palatini gravity, the symmetry of Weyl invariance can be implemented in a particularly simple way. Because the independent connection is inert under Weyl transformations, the scalar curvature carries weight \(-2\), while the measure carries weight \(+4\) in four dimensions. This makes \(\sqrt{-g}\,\RPal^2\) exactly Weyl invariant~\cite{Edery:2019txq}. More precisely, local scale invariance uniquely selects \(f(\mathcal{R})\propto\mathcal{R}^2\) off shell when restricting to Palatini \(f(\mathcal{R})\) models. Moreover, on shell and under a restricted set of conditions, \(\mathcal{R}^2\) is the unique non-topological representative of the ultraviolet Weyl-invariant sector. The ultraviolet endpoint of AWIG need not be put in by hand; it may arise naturally as the unique endpoint compatible with the symmetry assumption. Whether this endpoint is dynamically realised as a quantum renormalization-group fixed point remains an open question.

This uniqueness feature helps clarify the role of the curvature-dependent exponent \(n(\mathcal{R}_{*})\). The exponent parametrises how the curvature-scaling law interpolates between its two endpoint forms. As specified in the Introduction, the dimensionless variable \(\mathcal{R}_{*}\) is used throughout whenever curvature appears as the argument of the exponent or of the effective flow. This curvature dependence should not, however, be confused with a Wilsonian coarse-graining scale.

The curvature-flow analysis provides a complementary characterisation of one assumed class of smooth autonomous interpolations. Requiring the endpoint values \(n=1\) and \(n=2\), together with a genuine approach to the homogeneous \(\RPal^2\) endpoint, constrains the analytic structure of the effective flow within that class. The ultraviolet condition \(T(\mathcal{R})\to0\) forces the zero at \(n=2\) to be simple and requires \(p(2)<-2\). The infrared regularity and nondegeneracy condition likewise forces a simple zero at \(n=1\). This endpoint behaviour motivates a family of interpolating functions for \(n(\mathcal{R}_{*})\), whose infrared form agrees with the expression independently derived via the spectral dimension in Ref.~\cite{Coumbe:2025ktl}, modulo a derivative term. The particular interpolation is not unique, and the present analysis is not a microscopic renormalization-group derivation.

The main technical result concerns the ultraviolet counterterm structure of the strict \(n(\mathcal{R}_{*})\to2\) limit. On the regular branch \(\mathcal{R}\neq0\), and within the restricted theory space considered here, the equations of motion collapse all admissible on-shell divergences to the original \(\sqrt{-g}\mathcal{R}^2\) term plus the Euler density. This yields conditional on-shell closure of the ultraviolet sector at arbitrary loop order and therefore nontrivial evidence for improved ultraviolet behaviour relative to Einstein gravity, since in Einstein gravity new non-renormalizable counterterm structures proliferate at each loop order~\cite{tHooft:1974bx,Goroff:1985th}.

There may also be a broader significance to AWIG. As discussed in the Introduction, the causal structure theorems of Malament~\cite{Malament1977} and of Hawking et al.~\cite{Hawking:1976fe} suggest that much of spacetime geometry is already encoded in its causal structure, with the conformal factor being the principal obstruction to completeness. AWIG provides a concrete framework in which that obstruction---absolute local scale---is removed asymptotically. In this sense, AWIG may be viewed as a plausible step toward a complete background-independent theory of gravity.

Several questions, however, remain open. One is to test the robustness of AWIG by enlarging the theory space beyond the minimal Palatini \(f(\RPal)\) sector to include derivative-dependent exponents. Such terms are motivated both by the dimensional-reduction analysis in Ref.~\cite{Coumbe:2025ktl}, which suggests that the exponent \(n\) may contain differential structure, and by the possibility that a genuinely differential relation could soften some of the known pathologies of Palatini models~\cite{Olmo:2011uz,Sotiriou:2008rp}. Their implications for strong-field phenomenology and ultraviolet closure, however, require a separate analysis and are left for future work. At the quantum level, it would be important to go beyond the on-shell analysis given here, study possible anomalies, and determine whether the symmetry-based closure of the ultraviolet sector survives in a more complete treatment. A functional-renormalization-group calculation with independent metric and connection variables would be a natural way to test whether the phenomenological curvature flow used in Sec.~3 is generated dynamically and whether additional operators enter the quantum theory space. A further question concerns matter coupling. In the Palatini formulation, minimally coupled matter fields generically modify the connection equation of motion through the stress-energy trace, introducing algebraic constraints on matter that are absent in metric \(f(\mathcal{R})\) theories~\cite{Olmo:2011uz}. Whether these features persist in AWIG or are ameliorated by the curvature-dependent exponent \(n(\mathcal{R}_{*})\) requires a separate analysis. The ultraviolet closure results of Section~4 apply only to pure gravity.

The main conclusion may now be stated simply. If one takes seriously the idea that a final theory of gravity should not depend fundamentally on absolute local scale, then in four dimensions the Palatini \(\RPal^2\) theory may be singled out as unique, and AWIG is the corresponding interpolation between that ultraviolet endpoint and the infrared Einstein-like regime. This framework is conceptually well motivated, mathematically coherent, and exhibits nontrivial evidence for improved ultraviolet behaviour relative to Einstein gravity, beyond its already established power-counting renormalizability~\cite{Coumbe:2025ktl}. For these reasons, AWIG stands as a well-motivated and mathematically tractable candidate for a symmetry-driven ultraviolet extension of general relativity.

\section*{Acknowledgments}

The author is very grateful to Aria Rahmaty for useful early discussions related to this work. An AI-based tool was used to check and refine the proof of Proposition~4.1. All results and conclusions were independently verified by the author.

\bibliographystyle{unsrt}
\bibliography{references}

\end{document}